\documentstyle[aaspp4]{article}

\slugcomment{Submitted to Astrophys. J. Letters}

\lefthead{Pre\'s and Phillips}
\righthead{MAGNETIC ASSOCIATION OF CORONAL BRIGHT POINTS}
\begin{document}

\title{THE MAGNETIC ASSOCIATION OF CORONAL BRIGHT POINTS }
\author{Pawe{\l} Pre\'s\altaffilmark{1} and Kenneth J.H. Phillips}
\affil{Space Science Department, Rutherford Appleton Laboratory, Chilton, 
Didcot, Oxon OX11 0QX, U.K.}

\altaffiltext{1}{Royal Society/NATO Postdoctoral Fellow, also at Astronomical Institute, Wroc{\l}aw University, Kopernika 11,
51-622 Wroc{\l}aw, Poland, pres@astro.uni.wroc.pl}

\begin{abstract}

Several coronal bright points identified in a quiet region of the Sun  from
{\em SOHO}/EIT \ion{Fe}{12}  images are found to have a time evolution from
birth to decay  which is extremely well correlated with the rise and fall of
magnetic flux as determined from photospheric  magnetograms made with the  {\em
SOHO}/MDI instrument. Radiative losses of the bright points are found to be
much less than conductive losses but the sum of the two is comparable with the
available energy of the associated magnetic field. Several small `network
flares' occur during the lifetimes of these bright points, emphasizing the
strong connection of such events with the magnetic field.

\end{abstract}


\section{Introduction }

New high-resolution observations from the {\it Solar and Heliospheric
Observatory} ({\it SOHO}) are enabling comparisons to be made of extreme
ultraviolet (EUV) features in the solar atmosphere with the photospheric
magnetic field.  The association of coronal X-ray bright points (XBPs) with
ephemeral bipolar magnetic regions is well known (\cite{gol77}), but the
connection can now be more satisfactorily examined with data from the Michelson
Doppler Imager (MDI) on {\it SOHO} (\cite{sch95}), which forms photospheric
magnetograms with spatial resolution of 1.2$''$ at a frequency of approximately
one per 3 s, as well as coronal images from the {\it SOHO} Extreme Ultraviolet
Imaging Telescope (EIT) ({\cite{dela95}}) and the Soft X-ray Telescope (SXT)
(\cite{tsu91}) on the {\it Yohkoh} spacecraft.  Some results already discussed
({\cite{sch97}}; {{\cite{sch98}}) show that the photospheric magnetic field
outside of active regions is in a continuous state of replenishment, with flux
emerging from below and disappearing when collisions of flux concentrations
with opposite polarity occur. It is evident from MDI magnetograms how flux
concentrations break into fragments which then collide with other flux
concentrations of opposite polarity (resulting in partial or total flux
cancellation) or the same polarity (resulting in newly merged concentrations).

The results reported here are from a study of coronal bright points, 
identified from EIT \ion{Fe}{12} and {\em Yohkoh} X-ray images, and the
photospheric magnetic field  from MDI data, over a period of two days when
solar activity was at an  extremely low level. We examine the connection
between photospheric magnetic  field flux and the EUV flux and the occurrence
of several small `network  flares'. The estimated energy of the magnetic field
is compared with the energy that the coronal bright point plasma  loses by
conduction and radiation.

\section{Observations}

The period chosen for this study was 1997 July~31 (00:00 U.T.)  to August~2
(00:00 U.T.)  when solar activity was at almost the lowest level during the
recent solar minimum.  Two small active regions (NOAA 8064 and 8066) were on
the disk but showed no flaring activity, while the X-ray emission as detected
by the {\it GOES} monitoring satellites showed only tiny fluctuations with
steady component at the A1 ($1\times 10^{-8}$ W m$^{-2}$) level.  The Soft
X-ray Telescope (SXT) on {\em Yohkoh} performed $128\times 128$ pixel partial
frame rasters, covering the quiet region at the center of the Sun, with
half-resolution, i.e.  with pixel size 4.9$''$, so with a field of view
$10'\times 10'$.  The total observing time of this region was 4.5 hours with a
mean cadence of 25 s.  For most of the period, the thin aluminum (Al 0.1) and
aluminum/magnesium (Al/Mg) filters were used, with bandpasses in the
3--20~\AA\ range.

The EIT instrument regularly obtains full-Sun images at four different 
wavelengths, but those of interest here are made at 9 to 20 minute intervals in
a 170--220~\AA-wide band centered on a strong \ion{Fe}{12} line at  195~\AA,
which has peak emissivity at electron temperature $T_e \sim 1.4$~MK. We also
used MDI full-disk images averaged over five-minute intervals  which are
available every 90 minutes approximately.  

We co-aligned the images for study by allowing for differential rotation  using
the procedure of Howard et al. (1990), with time set at 00:00 U.T.  on 1997
August~1. Regions of bright \ion{Fe}{12} emission and significant magnetic 
field change over periods of only a few hours. We show representative  images
from MDI (gray-scale) and EIT \ion{Fe}{12} emission (black contours) in 
Fig.~1. Each image is a square with side 400$''$ (approximately $290\,000$
km).

\section{Analysis and Discussion}

\subsection{Magnetic Flux and Coronal Emission}\label{Mag-Cor}

Inspection of the MDI magnetograms shows many changes in the form of
flux-concentration motions which give rise to diverging and converging bipoles
and disappearance of magnetic elements.  Coronal bright points are observed in
SXT and EIT images, with lifetimes of several hours, which are co-spatial and
simultaneous with changes in the MDI magnetograms, normally in the form of
either diverging or converging bipoles.  We studied six such regions, marked A
to F in Fig.~1.

Using the measured magnetic field strengths in the MDI magnetograms, we
calculated values of magnetic flux $\Phi$ (in Mx), equal to $\int \mid B_l \mid
dS$ where $B_l$ is the magnetic field strength (in Gauss) measured along the
line of sight and the integral is over a small rectangular area $\int dS$ which
includes the region of significant \ion{Fe}{12} emission.  This area varied
according to the region (e.g.  for region A it was $36''\times 30''$ and for
region B $40''\times 25''$).

We found that the \ion{Fe}{12} emission  within each area is highly
time-correlated with $\Phi$, with rise and fall accomplished in a period of
about a day.  This is illustrated for the case of region A in Fig.~2. Here, the
general rise and fall of the \ion{Fe}{12} emission and $\Phi$ are nearly
coincident (note  that the $\Phi$ curve is much less frequently sampled) as are
the two peaks at approximately 00:00 U.T. and 08:00 U.T. on August~1. The
\ion{Fe}{12} flux for the six regions is almost proportional to $\Phi$.
Best-fit power-law relations between the two, of the form $F_{\rm Fe\,XII} =
F_0\ (\Phi/10^{18})^{\gamma}$ DN s$^{-1}$, showed that $\gamma$ was close to
unity. (One DN or digital number corresponds roughly to 18 electrons
for the EIT instrument, and to 100 electrons for the {\em Yohkoh} SXT.) Values
of $F_0$ and  $\gamma$ are given in Table~1. The X-ray signal from the SXT
instrument for this period is always very weak, but it is clear from the
light-curves of each  region that the emission is also correlated with the rise
and fall of $\Phi$ (top panel of Fig.~2). A similar relation was found for
global, total magnetic flux (from vector magnetograms) and global X-ray
luminosity by Fisher et al. (1998); their value of $\gamma$ (1.19) is similar
to those in Table~1.

To estimate the energy contained by the magnetic field, we took the field
configuration to be in the form of a semicircular loop with footpoints marked
by the two monopoles.  The loop is assumed to have uniform cross section with
area equal to  the mean of the two monopole areas. This gives a volume $V'$ and
a magnetic energy $V' B_l^2 /8\pi$.  We found that, at the time when  the
local value of $\Phi_l$ reaches its maximum, the energy over  the assumed loop
volume was $\sim 10^{29}$ erg for all analyzed  regions. 

 From the measured SXT flux in the Al 0.1 and Al/Mg filters we can calculate
temperatures $T_e$ and emission measures $EM=\int N_e^2 dV$ (cm$^{-3}$) with
$N_e$ the electron density of each bright point, following Tsuneta et  al.
(1991). These  are well determined for regions B, C and F, but not for the
remaining bright points because of the weak SXT signal. Plasma temperatures do
not change significantly during the period of SXT observation except for region
C, where the temperature decreased from 1.7 MK to 0.8 MK.  From SXT images we
estimated the volume $V$ of each bright point making the simple assumption that
emitting volume was either spherical or ellipsoidal, depending on its
appearance. (Note that this volume is in general larger than volume $V'$
mentioned above.) We then calculated the plasma thermal energy $3N_ekT_e V$.
These values, listed in Table~1, are about two orders of magnitude less than
the energy contained by the magnetic field.

Table~1 also compares bright point energy losses during  their lifetime due to
radiation and thermal conduction. Because SXT data cover  only 10\% of the
period of analysis and suffer discontinuities  due to orbital nights, we
calculated radiative losses by integrating the EIT  \ion{Fe}{12} light curve
scaled to the SXT fluxes. Typically, the radiative  energy loss is $\sim
10^{28}$ erg during the bright point lifetime, i.e.  only  10\%  of the
magnetic energy. For conductive energy losses we assumed Spitzer conductivity,
i.e. that energy is conducted down to the chromosphere  along field lines which
are taken to have simple geometry. As can be seen, conductive energy losses are
at least a factor ten more effective than radiative losses  and the total
energy losses are within a factor of 2 of the calculated magnetic energy.

\subsection{Network Flaring}\label{Flares}

Many short-lived enhancements can be identified in the  EIT \ion{Fe}{12} flux,
often consisting of a single point (the time cadence of these observations is
20 minutes). Such an enhancement occurs, e.g., in the general rise of the
emission of region A at around 22:13 U.T. on July~31 (see Fig.~2). Similar
enhancements were found by Krucker \& Benz (1998) in EIT data. We examined the
SXT flux for the appropriate interval for X-ray counterparts of these
flare-like events.  The signal is weak but we found at least  six significant
enhancements: Fig.~3 shows light curves of three of them. The time development
is not always fully observed because of the periodic  night-time intervals of
{\em Yohkoh}, but it is evident that there is a large variety of rise and decay
times. The X-ray equivalent of the \ion{Fe}{12}  enhancement in region B, for
example, is a very impulsive event with total duration of only five minutes,
whereas the event whose rise phase only is observed from region C at about
21:30~U.T. on July~31 is clearly much more  gradual. 

Using the ratio of fluxes in the SXT Al 0.1 and Al/Mg filters, we derived peak
temperatures and emission measures. These values, which are corrected by
subtraction of pre-event emission, are listed in Table~2. The temperatures are
between 1~MK and 2~MK while the emission measures are approximately $10^{45}$
cm$^{-3}$, or about $10^{-5}$ times smaller than a very large solar flare. As
before, the emitting volumes were estimated from the SXT images, enabling the
total thermal energies $E_{\rm th}$ to be determined.  These and the values of
$N_e$ are given in Table~2. Also, total emitted energies are given, estimated
from the temperature and the radiative loss data of Mewe et al. (1985).

These tiny events are very likely the `network flares' noted  by Krucker et al.
(1997) from SXT and centimeter-wavelength radio data. The flares discussed by
them have very similar characteristics to those listed in Table~2, and may be
classed as micro-flares in view of their thermal energies and emission measures
compared with very large solar flares. What we have established here is that
they are spatially coincident with magnetic features identifiable from the MDI
magnetograms. They occur during various stages in the time evolution of their
associated coronal bright points, from development to  decay. There is also no
apparent preference for the type of magnetic activity, with network flares
being emitted by emerging flux regions like F as well as cancelling flux
regions like C.  It appears, then, that network flares have very similar
properties to conventional, active-region flares (association with bipolar
magnetic field, large variety of time developments in soft X-rays, impulsive
microwave burst preceding the soft X-ray emission); only their size and
location (quiet-Sun regions) distinguish them. 

The frequency of the flares averaged over the whole solar surface can be
estimated from our data, but with only 6 events our result will be subject to
at least a 40\% uncertainty. The  six flares occurred during 4.5 hours of
effective SXT observation time over an area of $400''\times 400''$, which
corresponds to about $2300\pm 1000$ per day over the whole Sun.   The average
thermal energy of our flares is $7\times 10^{26}$ erg (Table~2), so the energy
rate requirement is $2\times 10^{25}$ erg s$^{-1}$, which is exactly the same
as the corresponding rate of Krucker et al. (1997). Because these events were
identified by eye inspection, the above values should be treated as lower
limits only.

\section{Summary and conclusions}

We have established that the magnetic flux $\Phi$ as determined from  the
photospheric magnetic field is highly time-correlated with the \ion{Fe}{12}
ultraviolet line emission from the six coronal bright points studied here.
There appears to be a good time correlation also with the X-ray flux though the
X-ray emission is in all cases very weak. For three of the bright points for
which temperature and emission measure are reasonably well determined, the
amount of magnetic energy, calculated from the simple model loop indicated in
section~\ref{Mag-Cor}, is within a factor of two of the total energy loss of
the plasma over its lifetime.  In view of the uncertainties of the magnetic
energy calculation and the determination of the energy losses, we consider that
the difference  between magnetic energy and total energy loss is insignificant,
i.e. that the magnetic energy supplies practically all the energy lost by the
bright point plasma. 

This result may be considered in the context of  the ideas developed by
Falconer et al. (1997, 1998) from observations that  coronal features can be 
categorized into low-lying, ``core" loops and overlying, extended features,
each crossing neutral lines and possibly sheared. Core-field activity is found
to drive up to apparently 10 times more heating in the surrounding extended
loop configurations. They also found that magnetic flux cancellation in the
core region  occurs as a coronal heating agent in addition to the observed
field shear, true for both active regions and the quiet Sun. Thus, this core
field activity possibly drives most of the heating of the quiet-Sun corona. 

Our findings would not seem to support Falconer et al. if indeed the amount of
magnetic energy calculated from photospheric magnetograms is equal to that 
lost by a coronal bright point in its lifetime by radiation and conduction.
Although there are factor-of-two uncertainties in the values of Table~1, it
appears that there is very little energy available from the magnetic field over
that needed to maintain the energy losses of the core configuration assuming
this is identifiable as the coronal bright point. 

We are, however, hesitant in asserting that there is a contradiction with the
findings of Falconer et al. (1997, 1998). The energy losses of the coronal
bright points given in Table~1 are almost entirely due to conduction, and they
are calculated on the assumption of  Spitzer conductivity acting along the
field lines with simple geometry.  However, the field geometry may be far from
simple, as shown by the presence of flare loop-top sources (Feldman et al.,
1994), and may involve turbulent fields as in the model of Jakimiec et al.
(1998). If complex field geometries applied to coronal bright points,  the
energy loss would be much less than calculated in section~\ref{Mag-Cor}. A
large fraction of the magnetic energy would then be available for heating of
the extended loop structures of the general corona, compatible with Falconer et
al. (1998).

Finally, we note the presence of several short-lived X-ray enhancements  which
are like very small flares and seem to be a common feature of coronal
bright points.  A lower limit to the energy rate over the whole Sun is 
very roughly $2\times 10^{25}$ erg s$^{-1}$.

\acknowledgments
P.P. acknowledges financial support through the U.K. Royal Society/NATO
Postdoctoral Fellowship Program. We thank the {\em SOHO} team for use of the
data used in this paper, notably the MDI team (P.I. P. H. Scherrer) and the EIT
team (P.I. J.-P. Delaboudini\`ere), and the {\em Yohkoh} team for use of
SXT images. Hugh Hudson is thanked for helpful discussions.

\begin{deluxetable}{crrcccc}
\footnotesize
\tablecaption{Power-law parameters for relation between observed EIT \ion{Fe}{12} flux (DN\,s$^{-1}$) and local magnetic flux $\Phi_l$,
$F_{\rm Fe\,XII} = F_0\ (\Phi/10^{18})^\gamma$ and comparison 
energies of coronal bright points (time of maximum  development).}
\tablewidth{0pt}
\tablehead{
\colhead{Region} & \colhead{$F_0$} & \colhead{$\gamma$} & \colhead{Thermal 
energy}  & 
\colhead{Radiative energy} & \colhead{Conductive energy} & \colhead{Magnetic 
energy}\nl
&&& \colhead{[ergs]} & \colhead{loss (in 1--300 \AA) [ergs]}\ & \colhead{loss 
[ergs]} & \colhead{[ergs]}
}
\startdata
A\tablenotemark{a}& 42 & 1.20 &  & & & $1.0\times 10^{29}$\nl
B& 38 & 1.04 & $3.8\times 10^{26}$ & $5.6\times 10^{27}$&$>4\times 10^{28}$ & 
$1.4\times 10^{29}$\nl
C& 160& 0.72 & $9.5\times 10^{26}$ & $1.1\times 10^{28}$&$1.4\times 10^{29}$ & 
$7.9\times 10^{28}$\nl
D\tablenotemark{a}& 41 & 1.07 &  & & & $ >2\times 10^{28}$\nl
E\tablenotemark{a,b}& 0.4& 1.95 & & & & $1.7\times 10^{29}$\nl
F& 86& 0.90  & $2.0\times 10^{27}$ & $6.0\times 10^{27}$&$2.6\times 10^{29}$ & 
$2.4\times 10^{29}$\nl
\enddata
\tablenotetext{a}{Plasma energy content and energy losses for regions A, D
and E are uncertain}
\tablenotetext{b}{Region E is located at the edge of the analysed area so
the estimated $F_{\rm Fe\,XII}$ and $\Phi_l$ may be uncertain.}
\end{deluxetable}

\begin{deluxetable}{crclllc}
\footnotesize
\tablecaption{Physical parameters of network flares from SXT observations}
\tablewidth{0pt}
\tablehead{
\colhead{Region} & \colhead{Time} & \colhead{$T_e$}   & \colhead{$EM$}     &
\colhead{$N_e$}     & \colhead{$E_{\rm th}$} & 
\colhead{Emitted energy} \nl
       &\colhead{(Date/U.T.)\tablenotemark{a}}    & 
\colhead{[MK]}&\colhead{[cm$^{-3}$]} &
       \colhead{[cm$^{-3}$]} & \colhead{[erg]}    & 
       \colhead{(in 1-300 \AA) [erg]} \nl 
}
\startdata
B & 1/08:42& 1.9 & $1.6\times 10^{45}$ & $1.6\times 10^9$ & 
$7.7\times10^{26}$& $1.3\times 10^{25}$ \nl
C & 31/21:45& 2.2 &$3.0\times 10^{44}$ & $1.3\times 10^9$ & 
$2.2\times10^{26}$& $>1 \times 10^{25}$ \nl
C & 1/00:37& 2.4 & $2.5\times 10^{45}$ & $2.0\times 10^9$ & 
$1.3\times10^{27}$& $5.5\times 10^{25}$ \nl
D & 1/05:27& 1.7 & $1.2\times 10^{45}$ & $1.8\times 10^9$ & 
$4.6\times10^{26}$& $5.0\times 10^{25}$ \nl
E & 1/08:43& 1.7 & $2.7\times 10^{45}$ & $2.7\times 10^9$ & 
$6.9\times10^{26}$& $6.3\times 10^{25}$ \nl
F & 31/23:25& 4.9 &$8.4\times 10^{43}$ & $5.9\times 10^8$ &
$2.9\times10^{26}$& $1.0\times 10^{24}$ \nl
\enddata
\tablenotetext{a}{`31/' means 1997 July 31, `1/' means 1997 August~1.}
\end{deluxetable}

\begin{figure}
\plotone{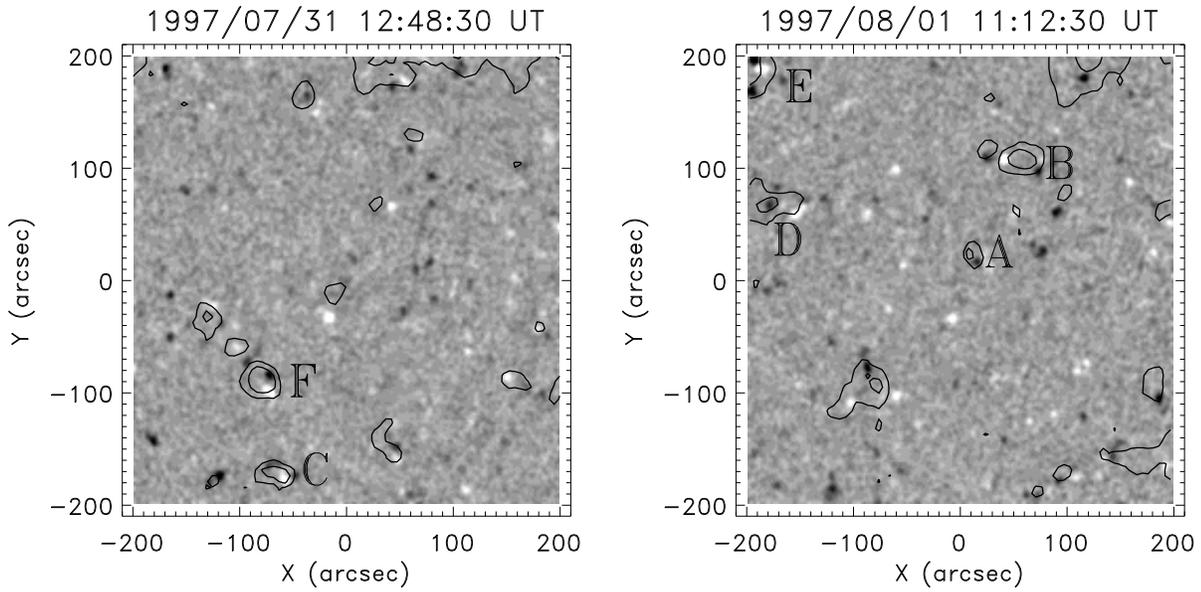}
\figcaption{
Two magnetograms of a quiet-Sun region at Sun center taken about a day apart,
during the period discussed in the paper. North is at the top, east on the left.
White areas correspond to positive polarity,  black to negative. 
Contours show \ion{Fe}{12} emission and correspond to levels of 120 and
200 DN/pixel/s.  Letters A to F denote the bright points discussed in the text.}
\label{Fe12_MDI}
\end{figure}

\begin{figure}
\plotone{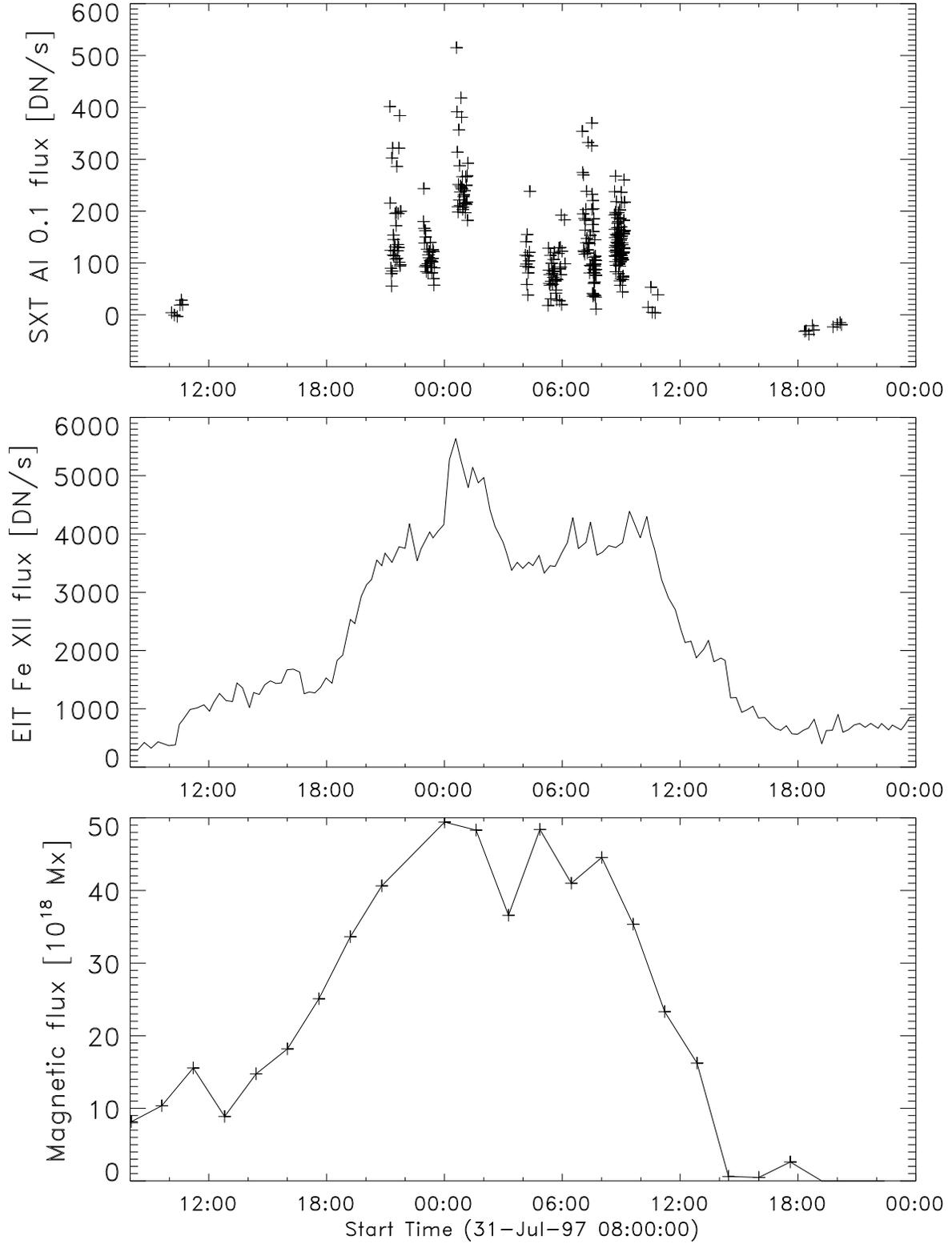}
\figcaption{The SXT Al 0.1 ({\em top}) and EIT \ion{Fe}{12} ({\em middle}) 
light-curves of the coronal bright point A, together with
({\em bottom}) time variations in the local magnetic flux $\Phi$ from the MDI instrument.}
\label{Fe12_MDI_lc}
\end{figure}

\begin{figure}
\plotone{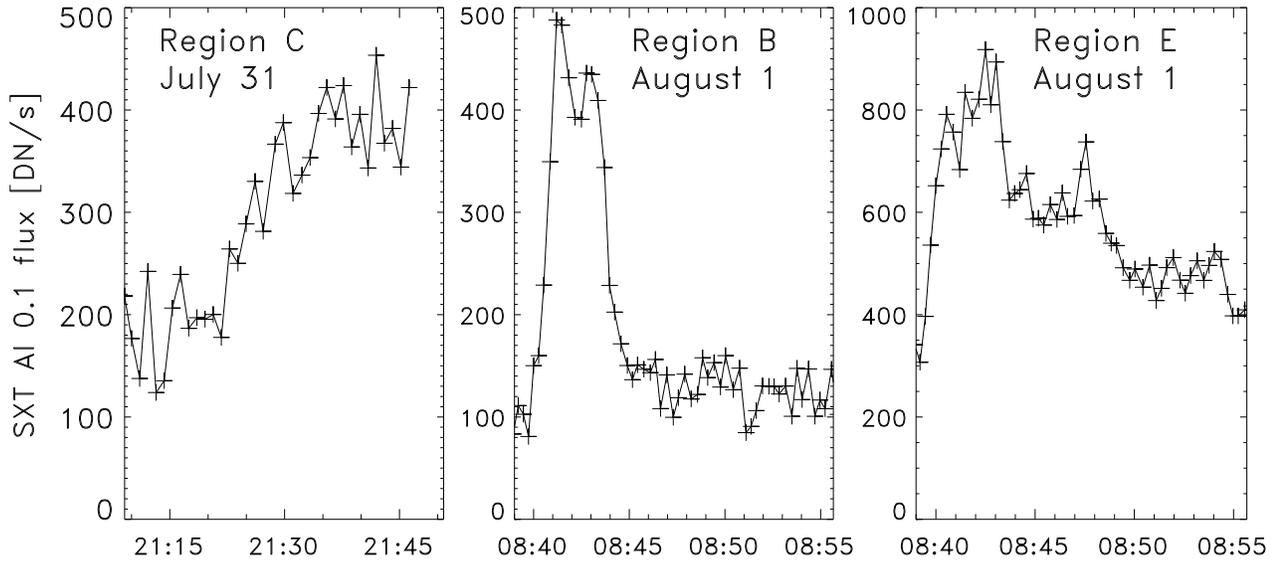}
\figcaption{Light-curves of three network flares observed by the {\em Yohkoh}
SXT instrument in the Al 0.1 filter.  
The locations of the flares are identified by the regions shown in Fig. 1.}
\label{SXT_flares}
\end{figure}


\begin{thebibliography}{}

\bibitem[Delaboudini\`ere et al., 1995]{dela95} Delaboudini\`ere, J.-P. 
et al. 1995, Sol. Phys., 162, 291
\bibitem[Falconer et al., 1997]{fal97} Falconer, D. A., Moore, R. L., Porter, 
J. G., Gary, G. A., \& Shimizu, T. 1997, \apj, 482, 519
\bibitem[Falconer et al., 1998]{fal98} Falconer, D. A., Moore, R. L., Porter, 
J. G., \& Hathaway, D. H. 1998, \apj, 501, 386 
\bibitem[Feldman et al., 1994]{fel94}Feldman, U., Hiei, E., Phillips, K. J. H.,
Brown, C. M., \& Lang, J., 1994, \apj, 421, 843
\bibitem[Fisher et al., 1998]{fis98} Fisher, G. H., Longcope, D. W., Metcalf, 
T. W., \& Pevtsov, A. A. 1998, \apj, submitted
\bibitem[Golub et al., 1977]{gol77} Golub, L., Krieger, A. S., Harvey, J. W., 
\& Vaiana, G. S. 1977, Sol. Phys., 53, 111
\bibitem[Howard, 1990]{how90} Howard, R.F., Harvey, J.W., \& Forgach, S. 1990,
Sol. Phys. 130, 295
\bibitem[Jakimiec et al., 1998]{jakim98}Jakimiec, J.,  Tomczak, M., Falewicz,
R., Phillips, K. J. H., \& Fludra, A., 1998. A\&A, 334, 1112
\bibitem[Krucker et al., 1997]{kru97} Krucker, S., Benz, A. O., Bastian, T. S., \& Acton, L. W. 1997, \apj, 488, 499
\bibitem[Krucker \& Benz, 1998]{kru98} Krucker, S., \& Benz, A.  O. 1998, \apj, 501, L213
\bibitem[Mewe et al., 1985]{mew85} Mewe, R., Gronenschild, E. H. B. M., \& 
van den Oord, G. H. J. 1985, \aaps, 62, 197
\bibitem[Scherrer et al., 1995]{sch95} Scherrer, P. H. et al. 1995, Sol. Phys., 
162, 129
\bibitem[Schrijver et al., 1997]{sch97} Schrijver, C. J., Title, A. M., 
van Ballegooijen, A. A., Hagenaar, H. J., \& Shine, R. A. 1997,\apj, 487, 424
\bibitem[Schrijver et al., 1998]{sch98} Schrijver, C. J. et al. 1998, Nature, 
394, 152
\bibitem[Tsuneta, 1991]{tsu91} Tsuneta, S. 1991, Sol. Phys., 136, 37

\end{thebibliography}
\end{document}